# Accelerating News Integration in Automatic Knowledge Extraction Ecosystems: an API-first Outlook

Juan M. Huerta   Clancy Childs

Dow Jones & Company
1211 Avenue of the Americas
New York, NY
Juan.Huerta@dowjones.com

## ABSTRACT
Leveraging Application Programming Interfaces (APIs) has been widely acknowledged as a valuable approach to software and system design that have promoted the acceleration of products and services development by allowing the decoupling of interface design from service implementation details. Many organizations in the news and journalism industry have adopted and promoted this API oriented approach. In the first part of this paper, we provide a survey of the most significant recent work around traditional news and journalistic open APIs and how these have been influenced by and impacted the news product landscape. In the second part of the paper, we identify two disruptive technology trends that we believe will impact the role and value of news/journalism products in the future: *API-first* development methodologies, and the increased role of news-supported automatic knowledge extraction and analytic services. We anticipate that these two driving forces will create a new wave of adoption, open collaboration, standardization and overall progress in news content adoption in knowledge platforms. We provide a brief overview of our experience in this area at Dow Jones.

## Categories and Subject Descriptors
D.2.2 [**Software Engineering**]: Design Tools and Techniques– *modules and techniques.*

## 1. Introduction
There has been substantial interest and momentum in the last decade around leveraging APIs across technology companies especially as a way for standardization, coordination, and widespread access to data and services [7]; news organizations have participated in this interest. So, What is an API? as Aitamurto et al. (2011) [10] quote from de Souza and Redmiles paper 2009 [5], *APIs are well-defined computing interfaces programmed to allow software programs to more readily "talk" to each other.* The main advantage of APIs is that they allow application and system developers, primarily, to carry out information hiding (Souza & Redmiles, 2009) which is an approach in which implementation details are hidden from other modules. The result is overall simpler code as well the ability to encapsulate best practices, hide complexity, and implement and with easier-to-adopt standards.

In terms of news and journalism, this interest has seen its manifestation more recently around both the development and publication of open APIs providing access to news and journalism content as well as the adoption of API-driven product development strategies in news organizations as part of broader open product technology efforts [10, 20]. These authors point out that the growth of news consumption across platforms, devices and operating systems, compounded by the increased demands for innovation (R&D) has pushed the news and journalism industry towards an open innovation model which early manifestation is the emergence of open APIs.

In this paper we analyze the state of journalistic APIs, and present an outlook on the direction that we feel APIs and journalism will follow in the future, particularly, how emerging technology trends are going to fundamentally affect the way APIs are leveraged in news organizations.

More specifically, in the first part of the paper we provide a brief survey of the state of APIs in the news and journalism industry and describe the adoption of open APIs to delivering content to external facing news applications and products that has taken place during recent years across several news organizations. We detail the reasons [10, 20] give on why leveraging API's in this way is intrinsically advantageous when applied in to news products: it addresses the challenges arising from the emergence of numerous platforms and channels, support of open services, et cetera as well as the need for product innovation in the way of open collaboration (Cheesbrough [9]). We also provide our perspective on the state of open collaboration.

In the second part of this paper we provide our outlook on the impact that two emerging technology trends will have in the future of news and journalism. These two trends are (*i*) emerging *API-first* development methodologies, and (*ii*) the maturity and progress reached by analytic and knowledge platforms. With respect to the first trend: the focus on APIs has recently been further extended to the *API-first* development methodology in which the specification of the API endpoints is carried out early in the design process as a way to provide an important architecture and system reference and design artifact. With respect to the second trend: we argue that news and journalism data and content is evolving from its historical role of supporting traditional user-facing products to a more general role of being source content to analytic and knowledge platforms. Together, we anticipate that API-first approaches will be of great use in accelerating product and process innovation as well as collaboration by (i) encapsulating the complexity of analytic and annotation engines and pipelines, (ii) abstracting the annotation and knowledge extraction process, and (iii) decoupling knowledge generation pipelines from client interfaces, platforms and applications. We detail the advantages that journalism can extract from formulating the process of news-to-knowledge distillation as well as the implications in the journalistic process in an API-driven context. We finalize this paper with a brief overview of our experience in this emerging area at Dow Jones and the implications we anticipate to the news and journalism industry.

## 2. APIs in News

In Journalism, APIs in general have further helped news organizations address multiple challenges arising from technology as well as from business realities: growing number of delivery platforms, media convergence resulting in increased competition, audience fragmentation by shifts in where people get their news [10, 34]. Additionally, and more recently, APIs have facilitated the implementation of more sophisticated solutions, including customer personalization at every touch point, acceleration of the news lifecycle, ability to rapidly support a fast rate of device innovation (watches, cell phones, tablets, etc.) with an ever-increasing product sophistication (notifications, personalized news, contextual news, etc.).

Open APIs refer to those technologies that support full interaction and communication across web sites through *public* interfaces. Open APIs can be considered as a public access entry point to a possibly proprietary service or software through programmatic approaches. Several open APIs made available by large Internet organizations like Google, Facebook, Youtube etc., are described in Bodle (2011) [8]. The primary result of these open APIs has been the creation and adoption of applications (and websites) by a wide swath of organizations and developers [4, 8]. Many of these Open APIs have enabled developers integrate social aspects (Twitter, Facebook connections) to their applications.

Open APIs in journalism [10, 23] have been described as a way to promote innovation in news organizations, specifically open innovation [6, 9, 20]. These Open APIs primarily have been used to expose articles and content to developers and users. The result has been an emergence of multiple APIs and services including NPR, Quartz, Guardian, NY Times, etc. We provide an overview of some of these in section 2.2

### 2.1 Best Practices

Best practices around designing and creating APIs have reached a significant level of maturity [23, 27, 33, 37, 38, 40] mostly as a result of open collaboration. We provide here a brief overview of some aspects of these best practices.

Current best practices include an API that represents a stateless, uniform interface, create consistent request/response patterns based on jsonapi.org specifications, maintain a shared Data Dictionary across the organization, specification of the extent of data caching, etc. Aspects include common tooling (e.g., swagger [40]), REST APIS ([21, 33]) as well as mechanisms to access to resources like media, hypermedia and annotations [30, 35, 36]. Specific best practices steps when designing a REST API: consists of selecting a protocol, identifying the resources, and finally mapping the actions. It's clear how, in the *news* model, resources naturally correspond articles (we will see how algorithms will play this role as well). API should also present further support for functions and best practices (patterns) for persistent storage (CRUD operations: Create, Retrieve, Update, Delete) and patterns for retrieving (including get, filter) for creating (posting), updating and deleting content [2].

For example, figure 1 below, shows a partial example of a GET and POST actions of a specification in JSON where the resource is newsletter, the actions are POST, query parameters are related to pagination and the response are paginated newsletters.

New roles of APIs as means to specifying collaboration points in distributed task protocols are emerging. An example of this can be found in [33].

```
Request 20 records starting from the 11th record
GET /newsletters?page[offset]=11&page[limit]=20

Note:
POST - post should be reserved for complex transactions
"paging": {
        "offset": 11,
        "limit" : 20
}

Response:
{
  "links": {
    "self": "http://api.dowjones.com/pib/1.0/newsletters",
    "next":
"http://api.dowjones.com/pib/1.0/newsletters?page[offset]=20&page[limit]=10",
    "last": ""http://api.dowjones.com/pib/1.0/newsletters?page[offset]=210"
  },
  "data": [{
    "type": "newsletters",
    "id": "12345",
```

**Figure 1**: Example of paginated content JSON specification for newsletter resource.

### 2.2 News & Journalism Public APIs

Several organizations have made available their open APIs in the last years. These APIs have played a role primarily a gateway to data and content. We now provide a brief overview of a few differentiated news and journalism APIs:

- The Guardian was one of the first organizations to introduce a public API, and today they provide access to over 1 million articles through open-platform.theguardian.com. Access and control actions to Hypermedia content is possible through their APIs [30, 35, 36, 37].

- The New York Times: introduced articles API (around 2009) and currently provides a very rich set of APIs providing access to various resources including semantic API, top stories API, and movie reviews API, among others.

- NPR: one of the first, introduced around 2008 [3], provides access to content and services including Transcript API, and Station Finder API.

- USA Today: two open APIs around 2010 provides access to news articles going back to 2004.

- Getty Images offers an API that allows search for photos [41] allowing developers to embed content into their applications.

- The Wordpress API provides access points to the Content Management System supporting extensions to WordPress through plugins.

- Quartz: launched in 2012, Quartz provides news access and has been described [17, 39] as close as any other site to being view agnostic. quarts.com/api provides a minimalist API to content.

The above selected examples illustrate the diversity and richness of current APIs providing access not only to article content but to resources (photos) as well as specialized content (movie reviews, etc.).

### 2.3 Open Innovation in News & Journalism

How well, in hindsight, have the current APIs address the Aitamurto's [20] requirements across news organizations? We can see from last section's list, that the availability of the current APIs clearly address the growth in devices, platforms and operating systems. Furthermore, we can ask: has the open collaboration model [6, 9, 23] taken place and succeeded? We see that two models have taken place: internal open collaboration (catalyst) and external open collaboration. Observed in [10], large organizations that adopted open APIs experienced niche product acceleration, development of broader product portfolios, as well as collaboration with external groups (e.g., iPhone app with NPR). This has been further confirmed Boyd's observation [35, 36] "the often-hidden growth among APIs: those that are largely consumed within their own industry verticals, the increasing use of private and partner APIs, and those open APIs used in specific communities".

Overall, these results of API as catalyst of innovation and collaboration are very encouraging. We anticipate further progress in the news and journalism will come from two driving factors, which we explain in the following 2 sections.

## 3. API-first in Content Systems

Much attention has been paid recently to the *API-first* development methodology in which, as mentioned earlier, the specification of the API endpoints is carried out early in the design process as a way to provide an important architecture and system reference and design artifact [26, 28].

In the case of news-driven content processing pipelines, where the primary resource is typically large collections news articles, as API-first approaches get adopted, attention will need to shift into abstracting content as *resources,* and abstracting analytics/transformation pipelines an *actions*. News organizations will also have to address the following points:

- *Extending CRUD to specific analytic pipelines and tasks*: As described in Section 2, a key step in designing REST APIs requires definition of the actions as well as resources. This means that the set of actions will have to evolve from simple set of basic CRUD operations into more complex resource and data abstractions and transformations that fit the domain and the task.

- *Designing multilayered solutions, encapsulating pipeline complexity; commoditization of services*: current content process pipelines can be be very algorithmically complex. Certain services that contain enough complexity that task specialization will become more common. Organizations will expose their complex task through public APIs, ([22]), which in turn will eventually will lead to standardization and eventually possibly commoditization (e.g., as has been the case for example of machine translation, information extraction, etc.). In the news domain, emerging examples of this can be seen in the NY Times public APIs reflecting common information extraction patterns on news content. Designing complex processing pipelines based on simpler, commodity services, will become prevalent in the industry.

- *Identifying value in resources and pipelines and developing monetization models*: currently it is quite possible to monetize content (content-as-service), algorithms, computation (Amazon AWS), and combinations thereof. News organizations will have the opportunity to provide content under different innovative monetization models (for example, computation-time spent on analysis-ready corpora, volume of bytes streamed, etc.).

As news organizations address these issues, further API standardization, best practice adoption, and open collaboration will occur, not unlike the open innovation described in section 2. This time, however, it will be centered on enabling not the support of device and platform applications (as the open APIs did previously) but rather the complexity inherent to content processing pipelines that will feed knowledge ecosystems.

## 4. News APIs in Knowledge Ecosystems

In the previous section we argued that API-first design approaches will likely drive solution architects to address the specific questions and issues related to CRUD, service encapsulation, resource and action definitions etc. This collective analysis will work in increased collaboration as solutions become standardized.

In this section, we focus on the second aspect in our innovation outlook: the growing role that proprietary content, in general, and news content in particular, is playing in knowledge generating pipelines (and the growing emergence and adoption of these platforms) will put pressure on news organizations to structure the design of these complex services through API-first design strategies and on occasion, make these APIs open creating in turn more opportunities for new and more complex services.

Together, these two driving aspects will feed each other resulting in increased collaboration, innovation, and standards adoption.

Traditionally, news content has played an important role as source data for knowledge generation platforms (e.g., [1]). However, historically, the *majority* of the services and annotators in the news domain pertain to relatively narrow information extraction tasks (e.g., entity extraction) or domain-specific tasks (e.g., movie reviews, stock movements, etc.). For example, the NY Times public APIs provide a relatively narrow set of services. We anticipate that as the adoption of news content becomes more prominent in semantic, knowledge and reasoning platforms, the services and annotators will become broader and more sophisticated.

The evolution and maturity of knowledge processing platforms is currently evident in recent growth in activity around open knowledge communities and applications like DBPedia [32], semantic reasoning platforms [12, 25], automatic knowledge discovery [13] and the IBM Watson ecosystem [19, 43], in addition to many other open and proprietary platforms. These more complex platforms and ecosystems are driving the demand for more complex analytic services and content processing pipelines as well as proprietary content itself, and thus, creating incentives for organizations to open these to external access. A clear example of this trend was announced recently when Goldman Sachs announced it is making available their algorithms, data, and tools to clients [42]. As more news organizations provide open APIs to content and news related services, news content will play a larger role in the knowledge ecosystem.

### 4.1 APIs as Catalysts in Open Ecosystems

As knowledge processing platforms become more pervasive and as news organizations provide content to these platforms and ecosystems, the adoption of public and open APIs will yield the following benefits which in turn will further accelerate the complexity and pervasiveness of news systems:

- APIs will help organizations design and disseminate innovative sets of resource types (for example, [14]) and action; a rich new taxonomy of resource types will likely emerge (e.g., comprising standardized resources like news related entities, time series and streaming events, breaking news services, comments data, media, etc.)

- APIs will encapsulate algorithmic complexity hiding it from client applications, news organizations, and developers; specialized annotators will likely emerge from specialized organizations allowing news organizations to focus on content (e.g., filters for media, translation for text, transcription for audio streams, information extraction, etc.)
- APIs will facilitate the use of content in different domains promoting the emergence of multiple new related products, applications and even user experiences in addition to facilitating the integration of news content services in diverse types of platforms (e.g., finance, education, consulting, government and planning, business intelligence, geo-analytics, advertisement and marketing, etc.) and the integration of heterogeneous services (e.g., maps, weather, etc.)
- Support the emergence of design patterns and the encapsulation best practices; creating and following industry standards will allow for news organizations to borrow from open source and commoditized technology stacks, ecosystems and execute vendor-agnostic roadmaps.
- Support complex internal workflow tools, allowing for the mixture of manual work, with automatic processes as well as crowd sourcing.
- Enable news and journalism ecosystem as parts of other industries and ecosystems (e.g., ad targeting, customer and subscription, social platforms, linked open data (Marden [18]) creating new revenue and monetization streams.

## 4.2 News-centric Knowledge Workflows

In addition to enabling participation of news content services in *open* knowledge ecosystems, the encapsulation of content and analytic services using API's can significantly impact complex *internal* workflows. For example, as the life cycle of a news story gets more sophisticated and complex every day [11, 16, 29, 31] reliance on sophisticated tools, pipelines and workflows increases as a result. As organizations develop complex internal pipelines to aggregate, curate, validate, edit, and process in general content, they rely on specialized content management platforms (CMPs) and tools. As a natural extension of our analysis in the previous section, designing workflows for journalism in an API-first fashion requires clearly establishing separation and discretization of tasks, definition of data resources, among other requirements which produce as a result benefits that mirror those of the external API counterparts and will yield similar catalyst-related benefits. External and distributed workflows will also benefit from API driven approaches ([15, 24])

## 5. Lessons Learned from 3 Dow Jones APIs

To illustrate our experience on news and API-first, we now briefly provide a few illustrative examples of *internal* content related Data Science exploratory projects at Dow Jones which illustrate some of the points described in the previous section:

- *Relation Discovery Service*: This system serves as an annotator service that encapsulates a complex Machine Learning classifier. It operates on xml documents (articles) and provides an annotated version of the article with additional classification information tags. It is intended to be internally invoked by content processing pipelines and used in human-in-the loop annotation workflows as a hypothesis generation step. *The main lesson learned from this system is that a complex Machine Learning algorithm can be encapsulated as a service through an API.*
- *Entity Community/Graph Analysis Service*: This system provides graph algorithmic services for our graph database of entities and relations. It computes micro networks, macro networks, calculates communities and network centrality among other services. It encapsulates and abstract basic graph theoretical approaches like graph traversals and connected components. Designing the API first, allowed to focus on the user interface and to decouple the algorithmic complexities from the database and front-end implementation and design. *The main lesson learned from this system is that sometimes the algorithm-type should be exposed at the endpoint level (APIs not only can serve data, they serve algorithms).*
- *Dynamic Content Summarization Service*: This system creates an automatic summarization on a topic, entity, idea, or industry given an arbitrarily set of articles. *API-first has help us identify the right resource and action of this domain which in turn decoupled the user interface from the algorithmic implementations.* It has helped us also identify the key parameters and actions of this domain as well as the nature and structure of the resulting summary data structure. *It also helped us realize the importance of aligning user interface, summary structures, algorithmic steps, and underlying dataset and use the API as a powerful design compact*.

## 6. Conclusion

We believe that the first generation of API work in news organizations was the response to the challenges arising from the growing number of platforms, devices, operating systems, etc. The results of that first generation work was multiple organization setting forth innovative services, open collaboration, and to some extent best practices.

We believe that a second generation of API focus in news organizations is emerging now as the result of 2 driving forces: API-first approaches and the opportunity of news content to become part of the emerging knowledge services ecosystems. Open APIs will drive standardization, adoption, best standards and open collaboration. The eventual result of this new wave of focus on APIs will be the opportunity to unlock the full value of news and information content.

## 7. REFERENCES

[1] Tanev et al. (2008) "Real-time news event extraction for global crisis monitoring." *Natural Language and Information Systems*. Springer Berlin Heidelberg, 2008.

[2] Battle and Benson (2008) "Bridging the semantic Web and Web 2.0 with representational state transfer (REST)." Web Semantics: Science, Services and Agents on the World Wide Web 6.1.

[3] (2008) "NPR launches open API: New programming tool enables digital media users to integrate and share NPR news content," accessible via http://www.npr.org/about/press/2008/071708.API.html.

[4] De Souza et al. (2009) On the roles of APIs in the coordination of collaborative software development. Computer Supported Cooperative Work 18(5): 445–475.

[5] De Souza et al (2009b) On the roles of APIs in the coordination of collaborative software development. Computer Supported Cooperative Work.

[6] Enkel et al. (2009) Open R&D and open innovation: exploring the phenomenon. R&D Management 39(4).

[7] Jacobson, Brail and Woods (2011); API's: A Strategy Guide; O'Reilly.

[8] Bodle (2011) Regimes of sharing: Open APIs, interoperability, and Facebook. Information, Communication & Society 14(3).

[9] Chesbrough (2011) Open Services Innovation: Rethinking Your Business to Grow and Compete in a New Era. San Francisco, CA.

[10] Aitamurto et al. (2011). Open APIs and news organizations: A study of open innovation in online journalism. Paper presented at the International Symposium on Online Journalism, Austin, TX, April 1, 2011.

[11] (2011) http://www.adweek.com/fishbowlny/the-new-convoluted-life-cycle-of-a-newspaper-story/249508

[12] Wei, et al. (2012) "A comprehensive ontology for knowledge representation in the internet of things." Trust, Security and Privacy in Computing and Communications (TrustCom), 2012 IEEE 11th International Conference on. IEEE, 2012.

[13] Begoli, et al. (2012) "Design principles for effective knowledge discovery from big data." Software Architecture (WICSA) and European Conference on Software Architecture (ECSA), 2012.

[14] Mariani et al. (2012) "Self-Organising News Management: the Molecules of Knowledge Approach," *Self-Adaptive and Self-Organizing Systems Workshops (SASOW), 2012 IEEE Sixth International Conference on*, vol., no., pp.235,240, 10-14 Sept. 2012.

[15] Ananny (2012) "Press-Public collaboration as infrastructure: Tracing news organizations and programming publics in application programming interfaces." American Behavioral Scientist.

[16] Van Der Haak et al. (2012) The Future of Journalism: Networked Journalism. International Journal of Communication, v. 6, p. 16.

[17] (2012) http://www.niemanlab.org/2012/09/quartz-the-new-biz-news-site-is-a-technological-and-structural-innovator-with-only-a-few-hiccups/

[18] Marden et al. (2013). Linked open data for cultural heritage: evolution of an information technology. In *Proc. of the 31st ACM international conference on Design of communication* (SIGDOC '13). ACM.

[19] Gliozzo et al (2013). "Semantic Technologies in IBM Watson." *ACL 2013*.

[20] Aitamurto et al. (2013) "Open innovation in digital journalism: Examining the impact of Open APIs at four news organizations." new media & society 15, no. 2 (2013): 314-331.

[21] Richardson et al. (2013). RESTful Web APIs. O'Reilly, Sebastopol, CA.

[22] Tiwana, Amrit. (2013) Platform ecosystems: aligning architecture, governance, and strategy. Newnes.

[23] Aitamurto (2013) Balancing between open and closed: Co-creation in magazine journalism Digital Journalism, Taylor & Francis.

[24] Lewis et al. (2013) "Open source and journalism: Toward new frameworks for imagining news innovation." Media, Culture & Society 35, no. 5.

[25] Cheptsov et al. (2013) "Making Web-Scale Semantic Reasoning More Service-Oriented: The Large Knowledge Collider." *Web Information Systems Engineering–WISE 2011 and 2012 Workshops*. Springer Berlin Heidelberg, 2013.

[26] Rivero, et al. (2013) "MockAPI: an agile approach supporting API-first web application development." *Web Engineering*. Springer Berlin Heidelberg, 2013.

[27] Rivero, et al. (2014) "An Extensible, Model-Driven and End-User Centric Approach for API Building." *Web Engineering*. Springer International Publishing.

[28] (2014) http://apievangelist.com/2014/08/11/what-is-an-api-first-strategy-adding-some-dimensions-to-this-new-question/

[29] Lewis & Westlund (2014). Actors, actants, audiences, and activities in cross-media news work: A matrix and a research agenda. Digital Journalism.

[30] Sobocinski, (2014) Hypermedia APIs: the benefits of HATEOAS. Feb. 27 2014, Programmable Web.

[31] Schifferes et al., (2014) Identifying and Verifying News through Social Media Developing a user-centred tool for professional journalists . Digital Journalism. Special Issue: Future of Journalism: In an age of digital media and economic uncertainty . Volume 2, Issue 3.

[32] Lehmann et al. (2014) "DBpedia-a large-scale, multilingual knowledge base extracted from wikipedia." Semantic Web Journal 5.

[33] Wright (2014)"Restful annotation and efficient collaboration." LREC, 2014.

[34] (2105) http://www.niemanlab.org/2015/07/new-pew-data-more-americans-are-getting-news-on-facebook-and-twitter

[35] Boyd, (2015) How the Guardian is Approaching Hypermedia Based API Infrastructure, Apr. 27 2015, Programmable Web.

[36] Boyd, (2015b) API Strategy & Practice/ API Days Kicks off in Berlin. Apr. 24 2015, Programmable Web.

[37] (2015) The guardian open-platform, http://www.slideshare.net/openplatform

[38] (2015) JSONAPI at jsonapi.org

[39] (2015) Quartz http://www.niemanlab.org/2015/05/quartz-is-an-api-the-path-ahead-for-the-business-site-thats-reshaping-digital-news/

[40] Swagger http://swagger.io/

[41] https://api.gettyimages.com/swagger/ui/index.html

[42] (2015) http://www.wsj.com/articles/goldman-sachs-to-give-out-secret-sauce-on-trading-1439371800

[43] https://developer.ibm.com/watson/